\documentstyle[twoside,fleqn,espcrc2,epsfig]{article}
\newcommand{\ec}{\end{center}}

\newcommand{\AmS}{{\protect\the\textfont2
  A\kern-.1667em\lower.5ex\hbox{M}\kern-.125emS}}

\title{Extended HQEFT Lagrangian and currents 
\thanks{Work partially supported by CICYT under contract AEN99-0692}}
\author{F. Bert\'o$^{a}$ and M. A. Sanchis-Lozano$^{a,b}$\thanks{E-mail: mas@evalo1.ific.uv.es}
\vspace{0.4cm}\\
(a) Departamento de F\'{\i}sica Te\'orica \\
\vspace{0.1cm}
(b) Instituto de F\'{\i}sica Corpuscular (IFIC)\\
Centro Mixto Universitat de Val\`encia-CSIC \\
Dr. Moliner 50, E-46100 Burjassot, Valencia (Spain)}

\begin{document}
\begin{abstract}
From the tree-level heavy quark effective Lagrangian 
keeping particle-antiparticle mixed sectors we derive the vector 
current coupling to a hard gluonic field
allowing for heavy quark-antiquark pair annihilation and creation.
\end{abstract}
\vspace{0.1in}
\maketitle

\def\slash{\!\!\!\!/}
\def\slash{\!\!\!/}

\section{A COMPLETE TREE-LEVEL HQEFT LAGRANGIAN}
In this work we keep both heavy quark and heavy antiquark coupled sectors
in the HQEFT Lagrangian. To our knowledge, until \cite{mas99}
only the paper by Wu \cite{wu} in the literature actually dealt with 
both heavy
quark and antiquark fields altogether in the effective Lagrangian. In 
the present work, based on \cite{mas97,mas99}, we pursue this line
of investigation further, looking for more symmetric
expressions and a more transparent physical interpretation. \par

Following the standard reference \cite{neub} we
introduce the effective fields for a heavy quark bound inside a
hadron moving with (four-)velocity $v$, as  
\begin{equation}
h_v^{(+)}(x)\ =\ e^{imv{\cdot}x}\ \frac{1+v{\slash}}{2}\ Q^{(+)}(x)
\end{equation}
\begin{equation}
H_v^{(+)}(x)\ =\ e^{imv{\cdot}x}\ \frac{1-v{\slash}}{2}\ Q^{(+)}(x)
\end{equation} 
where $Q^{(+)}(x)$ stands for the (positive energy) fermionic field describing 
the heavy quark in the full theory; $h_v^{(+)}(x)$
and $H_v^{(+)}(x)$ represent the $\lq\lq$large'' and $\lq\lq$small''
 components of a classical spinor field respectively.\par
Similarly for a heavy antiquark,
\begin{equation}
h_v^{(-)}(x)\ =\ e^{-imv{\cdot}x}\ \frac{1-v{\slash}}{2}\ Q^{(-)}(x)
\end{equation}
\begin{equation}
H_v^{(-)}(x)\ =\ e^{-imv{\cdot}x}\ \frac{1+v{\slash}}{2}\ Q^{(-)}(x)
\end{equation}
where $Q^{(-)}(x)$ stands for the (negative energy) anti-fermionic 
field in the full theory. 
\par
Introducing the residual momentum $q$ for a heavy quark of total
momentum $p$ as $p=mv+q$, the on-shell condition can be written
as 
\begin{equation}
v{\cdot}q=-\frac{q^2}{2m}
\end{equation}
The underlying idea when introducing the residual momentum is 
that once removed the large mechanical momentum associated to the 
heavy-quark mass, only the low-energy modes remain in the effective 
theory \cite{capitani}.
This is the standard way to handle heavy quarks in singly heavy
hadrons according to HQET. However, one 
may conjecture about the possibility of performing such an
energy-momentum shift by introducing a center-of-mass residual
momentum for processes involving creation or annihilation
of heavy quark-antiquark pairs at tree-level.  Hence only
low energy modes of the fields (about the heavy quark mass) 
would be involved likely making meaningful our approach within the framework 
of an effective theory.
 
In sum, the main difference of this paper w.r.t. other standard works is 
that we are concerned with the existence of terms in the
transformed  Lagrangian mixing the large components of the 
heavy quark and heavy antiquark fields, 
i.e. $h_v^{({\pm})}{\Gamma}h_v^{({\mp})}$, 
where ${\Gamma}$ stands for a combination of Dirac gamma matrices
and covariant derivatives, 
instead of directly writing currents as bilinears 
mixing both particle and antiparticle sectors as in \cite{schuler,braaten}.

\def\slash{\!\!\!\!/}

The tree-level QCD Lagrangian is our point of departure: 
\begin{equation}
{\cal L}\ =\ \overline{Q}\ (i\overrightarrow{D{\slash}}-m)\ Q
\end{equation}
where
\begin{equation}
Q\ =\ Q^{(+)}+Q^{(-)}\ =
\end{equation}
\[ \ \ \ e^{-imv{\cdot}x}\biggl[h_v^{(+)}+H_v^{(+)}\biggr]
+e^{imv{\cdot}x}\biggl[h_v^{(-)}+H_v^{(-)}\biggr]
\]
and $D$ standing for the covariant derivative
\[
\overrightarrow{D}^{\mu}\ =\ \overrightarrow{\partial}^{\mu}\ -\ igT_aA_a^{\mu}
\]
with $T_a$ the generators of $SU(3)_c$. Substituting (7) in (8)
one easily arrives at
\begin{equation}
{\cal L}\ =\ {\cal L}^{(++)}\ +\ {\cal L}^{(--)}\ +\ {\cal L}^{(-+)}\ +\ 
{\cal L}^{(+-)}
\end{equation}
where we have explicitly splitted the Lagrangian into four different
pieces corresponding to the particle-particle, antiparticle-antiparticle
and both particle-antiparticle sectors. The former one has the form   
\[
{\cal L}^{(++)}=\overline{h}_v^{(+)}iv{\cdot}\overrightarrow{D}h_v^{(+)}-
\overline{H}_v^{(+)}(iv{\cdot}\overrightarrow{D}+2m)H_v^{(+)}
\]
\[ \ \ \ \ \ \ \ \ 
+\ \overline{h}_v^{(+)}i\overrightarrow{D{\slash}}_{\bot}H_v^{(+)}+\overline{H}_v^{(+)}i\overrightarrow{D{\slash}}_{\bot}h_v^{(+)} \]
corresponding to the usual HQET Lagrangian still containing
both $h_v^{(+)}$ and $H_v^{(+)}$  fields. We employ the common notation where
perpendicular indices are implied according to
\[ D^{\mu}_{\bot}\ =\ D_{\alpha}\ (g^{\mu\alpha}-v^{\mu}v^{\alpha}) \]
\par
Regarding the antiquark sector of the theory
\[  
{\cal L}^{(--)}=-\overline{h}_v^{(-)}iv{\cdot}\overrightarrow{D}
h_v^{(-)}+\overline{H}_v^{(-)}(iv{\cdot}\overrightarrow{D}-2m)H_v^{(-)}
\]
\[ \ \ \ \ \ \ \ \ 
+\ \overline{h}_v^{(-)}i\overrightarrow{D{\slash}}_{\bot}H_v^{(-)}+
\overline{H}_v^{(-)}i\overrightarrow{D{\slash}}_{\bot}h_v^{(-)}
\]
\par
The latter expressions, considered as quantum field Lagrangians
, do not afford tree-level heavy quark-antiquark pair
creation or annihilation processes stemming from the terms
mixing ${h}_v^{(\pm)}$ and ${H}_v^{(\pm)}$ fields since they contain
either annihilation and creation operators of heavy quarks
or annihilation and creation operators of heavy antiquarks
separately.\par
Nevertheless there are two extra pieces in the Lagrangian (8):
\[
{\cal L}^{(-+)}=e^{-i2mv{\cdot}x} {\times}
\]
\[ \ \ \ \ \ \ \ \ 
[\ \overline{H}_v^{(-)}iv{\cdot}\overrightarrow{D}h_v^{(+)}
-\overline{h}_v^{(-)}(iv{\cdot}\overrightarrow{D}+2m)H_v^{(+)}
\]
\[ \ \ \ \ \ \ \ \ 
+\ \overline{h}_v^{(-)}i\overrightarrow{D{\slash}}_{\bot}h_v^{(+)}+
\overline{H}_v^{(-)}i\overrightarrow{D{\slash}}_{\bot}H_v^{(+)}\ ] \nonumber 
\]
and
\[
{\cal L}^{(+-)}=e^{i2mv{\cdot}x}\ {\times}
\]
\[ \ \ \ \ \ \ \ \
[\ -\overline{H}_v^{(+)}iv{\cdot}\overrightarrow{D}h_v^{(-)}
+\overline{h}_v^{(+)}(iv{\cdot}\overrightarrow{D}-2m)H_v^{(-)}
\]
\[ \ \ \ \ \ \ \ \
+\ \overline{h}_v^{(+)}i\overrightarrow{D{\slash}}_{\bot}h_v^{(-)}+
\overline{H}_v^{(+)}i\overrightarrow{D{\slash}}_{\bot}H_v^{(-)}\ ] 
\]
where use was made of the orthogonality of the $h_v^{(\pm)}$ and 
$H_v^{(\pm)}$ fields.
As could be expected, there are indeed pieces mixing both 
quark and antiquark fields
leading to the possibility of annihilation/creation processes 
deriving directly from the HQET Lagrangian. After all the Lagrangian (8) is
still equivalent to full (tree-level) QCD 
\footnote{Superscript ${\lq\lq}(+)/(-)$" on the effective fields labels
the particle/antiparticle sector of the theory \cite{georgi}. Actually 
$\overline{h}_v^{(+)}\ (\overline{h}_v^{(-)})$ corresponds to
negative (positive) frequencies associated with 
creation (annihilation) operators of quarks (antiquarks).
In fact some extra ${\lq\lq}+/-$" signs should be added on the 
conjugate fields, 
i.e. $\overline{h}_v^{(+)-}$ and $\overline{h}_v^{(-)+}$, which
however will be omitted to shorten the notation.}. 

Let us note that, at first sight, one might think that the rapidly 
oscillating exponential would make both ${\cal L}^{(-+)}$ and
${\cal L}^{(+-)}$ pieces to vanish, once integrated over all velocities
according to the most general Lagrangian \cite{georgi}.
However, notice that actually this should not be the case for momenta of 
order $2mv$ of the gluonic field present in the covariant derivative. 
In fact, only such high-energy modes will survive, corresponding to the 
physical situation on which we are focusing, i.e. heavy quark-antiquark 
pair annihilation and creation.

The heavy quark-gluon coupling for an annihilation process can be read off
from the Lagrangian piece:
\begin{eqnarray}
{\cal L}_{coupling}^{(-+)} &  & =e^{-i2mv{\cdot}x}\ gT_aA_{\mu}^a\ {\times} \nonumber \\
& [ & \overline{H}_v^{(-)}\ v^{\mu}\ h_v^{(+)}\ \ \ \ \ (a) \nonumber \\
& - & \overline{h}_v^{(-)}\ v^{\mu}\ H_v^{(+)}\ \ \ \ \ (b) \nonumber \\
& + & \overline{h}_v^{(-)}\ \gamma_{\bot}^{\mu}\ h_v^{(+)}\ \ \ \ \ \ 
(c) \nonumber \\
& + & \overline{H}_v^{(-)}\ \gamma_{\bot}^{\mu}\ H_v^{(+)}\ ]\ \ (d)  
\end{eqnarray}

Next we want to eliminate the unwanted degrees of freedom associated to
the $\lq\lq$small'' components $H_v^{(\pm)}$ in $(a)$, $(b)$ and $(d)$. 
The piece (c) is the leading one in the above development.

\begin{figure}[htb]
\centerline{
\epsfig{figure=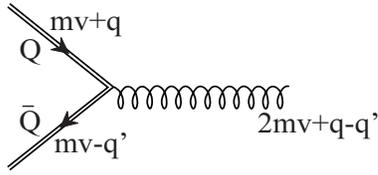,height=2.3cm,width=5.cm}}
\caption{Heavy quark-antiquark pair annihilation into
a gluon of momentum $2mv+q-q'$. In the center-of mass system
$q=(q_0,\bf{q})$ and $q'=(-q_0,{\bf q})$.
This process can be described meaningfully by HQEFT if
${\bf q^2}<m^2$, i.e. the square invariant mass of the gluon
is close to $4m^2$.}
\end{figure}

\def\slash{\!\!\!/}

\section{DERIVATION OF THE ANNIHILATION VERTEX}

Assuming the same conditions as in figure 1, we can write
for almost free heavy quarks in the initial- or final-state:
\begin{eqnarray}
H_v^{(+)} & = & (iv{\cdot}\overrightarrow{\partial}+2m-i\epsilon)^{-1}
i\overrightarrow{\partial{\slash}}_{\bot}h_v^{(+)} \\
H_v^{(-)} & = & (-iv{\cdot}\overrightarrow{\partial}+2m-i\epsilon)^{-1}
i\overrightarrow{\partial{\slash}}_{\bot}h_v^{(-)}
\end{eqnarray}
and for the conjugate fields
\begin{eqnarray}
\overline{H}_v^{(+)} & = & 
\overline{h}_v^{(+)}i\overleftarrow{\partial{\slash}}_{\bot}
(iv{\cdot}\overleftarrow{\partial}-2m+i\epsilon)^{-1} \\
\overline{H}_v^{(-)} & = & 
\overline{h}_v^{(-)}i\overleftarrow{\partial{\slash}}_{\bot}
(-iv{\cdot}\overleftarrow{\partial}-2m+i\epsilon)^{-1}
\end{eqnarray}

Let us note that we are using the free particle equations 
which can be viewed as derived from the non-interacting parts of
the Lagrangians ${\cal L}^{(++)}$ and ${\cal L}^{(--)}$
respectively. In fact, 
neglecting the soft gluon interaction among heavy quarks amounts to 
describe them as plane waves, i.e. actually no bound states
as a first approximation \cite{hussain}. Therefore assuming a
plane wave dependence of the field quantities, we can write
in momentum space
\begin{equation}
u(mv+q)=\biggl[\ 1+\frac{q{\slash}_{\bot}}{2m+v{\cdot}q}\ \biggr]\ u_h(q)
\end{equation}
\begin{equation}
\overline{v}(mv-q')=\overline{v}_h(-q')\ 
\biggl[\ 1-\frac{q{\slash}'_{\bot}}{2m-v{\cdot}q'}\ \biggr]
\end{equation}
where $u(p)$ ($v(p')$) denotes the full QCD spinor (antispinor) 
whereas $u_h(q)$ 
($v_h(-q')$) represents the HQET spinor (antispinor), i.e. obeying 
$v{\slash}u_h=u_h$  ($v{\slash}v_h=-v_h$).

Actually, in order to arrive to Eqs. (14-15) from Eqs. (10-13) one
has to expand the denominators as power series of derivatives
acting on the $x$-dependent factors, assumed exponentials,
to be finally resummed as a geometric series of ratio
 $v{\cdot}q/2m=-v{\cdot}q'/2m=-q^2/4m^2$ according to the on-shell condition
(5).
As a consequence, Eqs. (14-15) are only valid under the condition 
$-q^2<4m^2$, which implies ${\bf{q}^2}<8m^2$. Therefore, the
requirement ${\bf{q}^2}<m^2$ satisfies the above condition and
will allow a later non-relativistic expansion.

Substituting the above equations into the ${\cal L}_{coupling}^{(-+)}$
Lagrangian (9), we readily get for the on-shell heavy quark (vector) 
current coupling to a gluon, suppressing color indices and matrices

\[
\overline{v}_h\ \biggl[{\gamma}^{\mu}_{\bot}+
\frac{q{\slash}'_{\bot}-q{\slash}_{\bot}}{2m+v{\cdot}q}v^{\mu}
+\frac{q{\slash}'_{\bot}{\gamma}^{\mu}q{\slash}_{\bot}}
{(2m+v{\cdot}q)^2}\biggr]\ u_h 
\]
which can also be written as
\begin{equation}
\overline{v}_h\ \biggl[{\gamma}^{\mu}_{\bot}+
\frac{i{\sigma}^{\mu\nu}(q'_{\bot}-q_{\bot})_{\nu}}{2m+v{\cdot}q}
+\frac{q{\slash}'_{\bot}{\gamma}^{\mu}q{\slash}_{\bot}}
{(2m+v{\cdot}q)^2}\biggr]\ u_h 
\end{equation}
since \cite{mas99}
\[
P_-\ (v^{\mu}{\gamma}^{\nu})\ P_+\ 
=\ P_-\ (i{\sigma}^{\mu\nu}\ +\ 
{\gamma}^{\mu}v^{\nu})\ P_+
\]
with the projectors $P_{\pm}=(1{\pm}v{\slash})/2$, and 
$q'_{\bot\nu}v^{\nu}=q_{\bot\nu}v^{\nu}=0$.

\subsection{CENTER-OF-MASS FRAME}
\vskip 0.5 cm

We shall make use of the anticommutation relation:

\[ q{\slash}'_{\bot}{\gamma}^{\mu}q{\slash}_{\bot}=
2{q'}^{\mu}_{\bot}q{\slash}_{\bot}-
{\gamma}^{\mu}q{\slash}'_{\bot}q{\slash}_{\bot} \]

Now, in the center-of-mass frame we can write
$q{\slash}'_{\bot}q{\slash}_{\bot}=-{\bf q^2}$,
leading to the expression for the heavy quark vector current
\begin{equation}
\overline{v}_h\ \biggl[\frac{2E_q}{E_q+m}{\gamma}^{\mu}_{\bot}+
\frac{2{q'}^{\mu}_{\bot}q{\slash}_{\bot}}{(E_q+m)^2} \biggr]\ u_h
\end{equation}
where $E_q=m+q^0=\sqrt{m^2+{\bf q}^2}$. The following identity is 
satisfied (in the Dirac representation):

\begin{equation}
P_-\ q{\slash}_{\bot}\ P_+ = 
\left(
     \begin{array}{ll}
      0 & 0 \\
      {\bf \sigma}{\cdot}{\bf q} & 0     
     \end{array}
\right)
\end{equation}

Therefore identifying as two component spinors 
$u_h=\sqrt{E_q+m}\ \xi$, and 
$v_h=\sqrt{E_q+m}\ \eta$, we obtain from (17) the analogous expression
to Eq. (A.9b) of Ref. \cite{braaten}, 

\begin{equation}
{\eta}^{\dag}\ 
 \biggl[2E_q\ {\bf\bf {\sigma}}+
\frac{2{\bf q}\ {\bf {\sigma}}{\cdot}{\bf q}}{(E_q+m)} \biggr]\ {\xi}
\end{equation}
i.e. the vertex obtained from full QCD in terms of the Pauli spinors
$\xi$ and $\eta$,  
allowing a systematic non-relativistic expansion: the leading term
$2m(\eta^{\dag}{\bf \sigma}\xi)$ associated with (c) in (9) as
expected.

\section{SUMMARY AND LAST REMARKS}

We have derived a complete tree-level heavy-quark transformed Lagrangian in 
terms of the effective fields $h_v^{(\pm)}$, keeping the 
particle-antiparticle mixed pieces allowing for heavy quark-antiquark 
pair annihilation/creation. Let us note that such pieces are not
generally shown in similar developments in the literature.

Indeed, it may seem quite striking that a low-energy effective theory 
could be appropriate to deal
with hard processes such as $Q\overline{Q}$ annihilation or creation.
The keypoint is that assuming a kinematic regime where heavy quarks/antiquarks
are almost on-shell and moving with small relative velocity, 
the strong momentum dependence associated with the heavy
quark masses can be factored out, so that a description 
based on the low frequency
modes of the fields still makes sense. Such a kinematic regime is well 
matched by heavy quarkonium states and by colored
intermediate states in the so-called color-octet model recently
introduced \cite{fleming}  to account for high production rates of 
heavy quarkonia at the Fermilab Tevatron \cite{fermi}. \par

In particular, we have focused on an annihilation process with initial-state 
quarks satisfying the Dirac equation of motion for free fermions. Thus, 
we have derived directly from the ${\cal L}^{(-+)}$ piece the heavy quark
vector current coupling to a background gluonic field, recovering a well known
expression shown in the literature \cite{braaten}. A similar development can 
be done for a heavy quark pair creation process from the ${\cal L}^{(+-)}$
piece.  

The main point stressed in this paper is that one can derive the
heavy quark/antiquark vector current coupling to a hard gluonic field
from the corresponding piece of the transformed tree-level Lagrangian
(8), without resorting to an {\em ad hoc} construction from the
fermionic spinors themselves, perhaps providing a more self-consistent
basis to the matching procedure of the effective theory onto 
full QCD \cite{braaten}.

\end{document}